\documentclass[11pt,a4wide]{article}
\usepackage{amssymb}
\usepackage{amsmath}
\usepackage[english]{babel}
\usepackage{graphicx}
\usepackage{cite}
\usepackage{float}
\def\Journal#1#2#3#4{{#1} {\bf #2}, #3 (#4)}

\def\NIMA{{\em Nucl. Instrum. Methods} A}

\def\PLB{{\em Phys. Lett.}  B}
\def\PRL{\em Phys. Rev. Lett.}
\def\PRD{{\em Phys. Rev.} D}

\def\APJ{\em Astrophys. J.}

\def\IJMPA{{\em Int. J. Mod. Phys.}  A}

\def\BWP{\em Bled Workshops in Physics}
\def\s{{\,\rm s}}

\def\MeV{\,{\rm MeV}}

\def\({\left(}
\def\){\right)}
\def\cm{{\,\rm cm}}

\def\kpc{{\,\rm kpc}}
\def\beq{\begin{equation}}
\def\eeq{\end{equation}}
\def\bea{\begin{eqnarray}}
\def\eea{\end{eqnarray}}

\begin{document}
\begin{center}
        \large \textbf{Dark atoms and the positron-annihilation-line excess\\ in the
        galactic bulge}
    \end{center}

    \begin{center}
   J.-R. Cudell$^{1}$, M. Yu. Khlopov$^{2,3,4}$, and Q.Wallemacq$^{1}$
\end{center}
\begin{center}

    {\small \emph{$^{1}$IFPA, D\'ep. AGO, Universit\'e de Li\`ege, Sart Tilman, 4000
    Li\`ege, Belgium \\
    $^{2}$National Research Nuclear University ``Moscow Engineering Physics
    Institute'', 115409 Moscow, Russia \\
    $^{3}$ Centre for Cosmoparticle Physics ``Cosmion'', 115409 Moscow, Russia \\
$^{4}$ APC laboratory 10, rue Alice Domon et L\'eonie Duquet \\75205
Paris Cedex 13, France}}

    \end{center}

\medskip

\begin{abstract}
It was recently proposed that stable  particles of charge $-2$,
$O^{--}$, can exist and constitute dark matter after they
bind with primordial helium in O-helium (OHe) atoms.
We study here in details the possibility that this model
provides an explanation for the excess of gamma radiation in
the positron-annihilation line from the
galactic bulge observed by INTEGRAL. This explanation assumes that OHe, excited to a
$2s$ state
through collisions in the central part of the Galaxy, de-excites
to its
ground state via an E0 transition, emitting an electron-positron
pair.  The cross section for
OHe collisions with excitation to $2s$ level is calculated and it
is shown that the rate of such excitations in the galactic bulge
strongly depends not only on the mass of O-helium, which is
determined by the mass of $O^{--}$, but also on the density and
velocity distribution of dark matter.

Given the astrophysical
uncertainties on these distributions, this mechanism constrains
the $O^{--}$ mass to lie in two possible regions. One of these
is reachable
in the experimental searches for stable multicharged particles at
the LHC.

%\keywords{Elementary particles; dark matter; atomic collisions; positron annihilation.}
\end{abstract}
\section{Introduction}
According to modern cosmology, dark matter corresponds to
$25\%$ of the total cosmological density, is nonbaryonic and
consists of new stable particles. Such particles (see
\cite{book,Cosmoarcheology,Bled06,Bled07,newBook,DMRev} for reviews and references)
should
be stable, provide the measured dark-matter density and be decoupled
from plasma and radiation at least before the beginning of the matter-dominated era. It
was recently shown that heavy
stable particles of charge $-2$, $O^{--}$, bound to primordial helium in OHe atoms, can
provide an
interesting explanation for cosmological dark matter \cite{DMRev,DADM}. It should also
be noted that the nuclear cross section of the O-helium
interaction with matter escapes the severe constraints \cite{McGuire:2001qj,
McGuire2,ZF}
on strongly-interacting dark-matter particles
(SIMPs)\cite{McGuire:2001qj,McGuire2,ZF,Starkman,Wolfram,Starkman2,Javorsek,Mitra,Mack}
imposed by the XQC experiment\cite{XQC,XQC1}.

The hypothesis of composite O-helium dark matter,
first considered to provide a solution to the
puzzles of direct dark-matter searches, can
offer an explanation for another puzzle of modern astrophysics\cite{DMRev,DADM,I2}:
this
composite-dark-matter model can explain the excess of gamma radiation in the electron-
positron annihilation line, observed by INTEGRAL in the galactic bulge (see
\cite{integral} for a review and references). The explanation assumes that OHe provides
all the galactic dark matter and that its collisions in the central part of the Galaxy
result in 2s-level excitations of OHe which are de-excited to the ground state by an E0
transition, in which an electron-positron pair is emitted. If the 2s level is excited,
pair production dominates over the two-photon channel
in the de-excitation, because electrons are much lighter than helium nuclei, and
positron production is not accompanied by
a strong gamma-ray signal.

According to\cite{Finkbeiner:2007kk}
the rate of positron production $3 \cdot 10^{42}%S_3^{-2}
\s^{-1}$ is sufficient to explain the
excess in the positron annihilation line from the bulge measured by
INTEGRAL. In the present paper we study
the process of 2s-level excitation of
OHe from collisions in the galactic bulge and determine the conditions under which such
collisions can provide the observed excess.
Inelastic interactions of O-helium with matter in
interstellar space and subsequent de-excitation can give rise to radiation
in the range from a few keV to a few  MeV. In the galactic bulge with
radius $r_b \sim 1 \kpc$ the number density of O-helium can be of the order of
$n_o\approx 3 \cdot 10^{-3}/S_3 \cm^{-3}$ or larger, and the
collision rate of O-helium in this central region was estimated in
\cite{I2}: $dN/dt=n_o^2 \sigma v_h 4 \pi r_b^3 /3 \approx 3 \cdot
10^{42} S_3^{-2}
\s^{-1}$, with $S_3=m_{OHe}/1$ TeV. At the velocity of $v_h \sim 3 \cdot 10^7
\cm/\s$ energy transfer in such collisions is $\Delta E \sim 1 \MeV
S_3$. These collisions can lead to excitation of O-helium. If $OHe$
levels with nonzero angular momentum are excited, gamma lines should
be observed from transitions ($ n>m$) $E_{nm}= 1.598 \MeV (1/m^2
-1/n^2)$ (or from similar transitions corresponding to the case
$I_o = 1.287 \MeV $) at the level $3 \cdot 10^{-4}S_3^{-2}(\cm^2\ \s
\ \MeV$ ster$)^{-1}$.

\section{Collisional excitation cross section}

The studied reaction is the collision between two incident OHe atoms
in their ground states $1s$ giving rise to an OHe in an
excited $s$-state $ns$ while the other one remains in its ground
state : \begin{equation}
OHe(1s)+OHe(1s)\rightarrow OHe(1s)+OHe(ns)\label{eq:3}\end{equation}

If we work in the rest frame of the OHe that gets excited, and if we neglect its recoil after the collision, the differential cross section
of the process is given by \begin{eqnarray}
d\sigma\left(1s\rightarrow ns\right)=2\pi\left|\left\langle ns,\vec{p'}|U|1s,\vec{p}\right\rangle \right|^{2}\delta\left(\frac{p'^{2}}{2M}+E_{ns}-\frac{p^{2}}{2M}-E_{1s}\right)\frac{d^{3}p'}{\left(2\pi\right)^{3}}\label{eq:4}\end{eqnarray}
 where $M$ is the mass of OHe, $\vec{p}$, $\vec{p'}$ are the momenta
of the incident OHe before and after the collision, $E_{1s}$, $E_{ns}$
are the ground-state and excited-state energies of the target OHe and
$U$ is the interaction potential between the incident and the target
OHe's.

We shall neglect the internal structure of the incident OHe, so that its wave functions are plane waves.  $\psi_{\vec{p}}$ is normalized
to obtain a unit incident current density and the normalisation of $\psi_{\vec{p'}}$ is chosen for it to be pointlike, i.e. the Fourier transform of $\delta^{(3)}(\vec {r})$~\cite{Landau}:\begin{equation}
\begin{array}{cll}
\psi_{\vec{p}} & = & \sqrt{\frac{M}{p}}e^{i\vec{p}.\vec{r}}\\
\\\psi_{\vec{p'}} & = & e^{i\vec{p'}.\vec{r}}\end{array}\label{eq:5}\end{equation}
where $\vec{r}$ is the position vector of the incident OHe and $p=\left|\vec{p}\right|$.

In the following, we shall be lead to considering
$O^{--}$ masses which are much larger than the mass of helium or the bound state energies.  Therefore, the origin of the rest frame
of the target OHe coincides with the position of its $O^{--}$
component and its reduced mass $\mu$ can be taken as the mass of helium $ M_{He}$.

The OHe that gets excited is described as a hydrogenoid atom, with
energy levels $E_{ns}=-0.5M_{He}\left(Z_{He}Z_{O}\alpha\right)^{2}/n^{2}$
and initial and final bound-state wave functions $\psi_{1s}$, $\psi_{ns}$
of a hydrogenoid atom with a Bohr radius $a_{0}=\left({M_{He}Z_{He}Z_{O}\alpha}\right)^{-1}$.

The incident OHe interacts with the $O^{--}$ and helium components in the
target OHe, so that the interaction potential $U$ is the sum of the
two contributions $U_{O}$ and $U_{He}$: \begin{equation}
U\left(\vec{r}\right)=U_{O}\left(\vec{r}\right)+U_{He}\left(\vec{r}-\vec{r}_{He}\right)\label{eq:6}\end{equation}
 where $\vec{r}_{He}$ is the position vector of the helium component.

The first term $U_{O}$ gives a zero contribution to the integral of expression
\eqref{eq:4} since the states $\psi_{1s}$ and $\psi_{ns}$ are orthogonal. For the second term,
we treat the incident OHe
as a heavy neutron colliding on a helium nucleus through short-range nuclear forces.
The interaction potential can then be written in the form of a contact term:
\begin{equation}
U_{He}\left(\vec{r}-\vec{r}_{He}\right)=-\frac{2\pi}{M_{He}}a_{0}\delta\left(\vec{r}-\vec{r}_{He}\right),\label{eq:7}\end{equation}
where we have normalised the delta function to obtain an OHe-helium elastic cross section
equal to $4\pi a_{0}^{2}$.

Going to spherical coordinates for $\vec{p'}$ and integrating
over $p'=\left|\vec{p'}\right|$ in the differential cross section
\eqref{eq:4}, together with the previous expressions \eqref{eq:5},
\eqref{eq:6} and \eqref{eq:7}, we get \begin{equation}
d\sigma\left(1s\rightarrow ns\right)=\left(\frac{M}{M_{He}}\right)^{2}a_{0}^{2}\left(\frac{p'}{p}\right)\left|\int e^{-i\vec{q}.\vec{r}_{He}}\psi_{ns}^{*}\psi_{1s}d^{3}r_{He}\right|^{2}d\Omega\label{eq:8}\end{equation}
 where $\vec{q}=\vec{p'}-\vec{p}$ is the transferred momentum and
$d\Omega$ is the solid angle. From the integration over the delta
function in \eqref{eq:4}, we have obtained the conservation of energy
during the process:
\beq p'^{2}=p^{2}+2M\left(E_{1s}-E_{ns}\right).\label{eq:pp}\eeq It leads to the threshold energy corresponding to $p'^{2}=0$ and to
a minimum incident velocity $v_{min}=\sqrt{{2\left(E_{ns}-E_{1s}\right)/M}}$.
The previous expression for $p'$ allows us to express the squared
modulus of $\vec{q}$ as \begin{equation}q^{2}=2\left(p^{2}+M\left(E_{1s}-E_{ns}\right)-p\sqrt{p^{2}+2M\left(E_{1}-E_{ns}\right)\cos\theta}\right),\label{eq:q2}\end{equation}
where $\theta$ is the deviation angle of the incident OHe with respect
to the collision axis in the rest frame of the target OHe.

$e^{+} e^{-}$ pairs will be dominantly produced if OHe is excited to a $2s$ state, since the only de-excitation
channel is in this case from $2s$ to $1s$. As  $e^{+} e^{-}$
pair production is the only possible channel, the differential pair production cross section $d\sigma_{ee}$
is equal to the differential collisional excitation cross section.
By particularizing expression \eqref{eq:8} to the case $n=2$, one
finally gets \begin{equation}
\frac{d\sigma_{ee}}{d\cos\theta}=512^{2}\left(\frac{2\pi M^{2}}{M_{He}^{2}}\right)a_{0}^{6}\left(\frac{p'}{p}\right)\frac{q^{4}}{2\left(4a_{0}^{2}q^{2}+9\right)^{6}}\label{eq:9}\end{equation}

\section{The $e^{+} e^{-}$ pair-production rate in the galactic bulge}
The total $e^{+} e^{-}$ pair production rate in the galactic bulge
is given by \begin{equation}
{\left. dN\over dt\right|}_{ee}=\int_{V_{b}}\frac{\rho_{DM}^{2}\left(\vec{R}\right)}{M^{2}}\left\langle \sigma_{ee}v\right\rangle \left(\vec{R}\right)d\vec{R}\label{eq:10}\end{equation}
 where $V_{b}$ is the volume of the galactic bulge, which is a sphere
of radius $R_{b}=1.5$ kpc, $\rho_{DM}$ is the energy density distribution
of dark matter in the galactic halo and $\left\langle \sigma_{ee}v\right\rangle $
is the pair production cross section $\sigma_{ee}$ times relative
velocity $v$ averaged over the velocity distribution of dark matter
particles. The total pair-production cross section $\sigma_{ee}$
is obtained by integrating \eqref{eq:9} over the diffusion angle.
Its dependence on the relative velocity $v$ is contained in $p$, $p'$
and $q$ through $p=Mv$ and the expressions (\ref{eq:pp}) and (\ref{eq:q2}) of $p'$ and $q$ in terms of $p$.

We use a Burkert~\cite{burkert} flat, cored, dark matter density profile known to reproduce well the kinematics of disk systems in massive spiral galaxies and supported by recent simulations including supernova feedback and radiation pressure of massive stars~\cite{Maccio:2011eh} in response to the cuspy halo problem:
%We use a spherical Burkert density profile \cite{burkert} presenting a central core
%and known to reproduce well the effect on the dark matter profile
%from baryons that have collapsed to the center:
\begin{equation}
\rho_{DM}\left(R\right)=\rho_{0}\frac{R_{0}^{3}}{\left(R+R_{0}\right)\left(R^{2}+R_{0}^{2}\right)},\label{eq:11}\end{equation}
 where $R$ is the distance from the galactic center. The central
dark matter density $\rho_{0}$ is left as a free parameter and $R_{0}$
is determined by requiring that the local dark matter density at $R=R_{\odot}=8$
kpc is $\rho_{\odot}=0.3$ GeV/cm$^{3}$. The dark matter mass enclosed
in a sphere of radius $R$ is therefore given by \begin{equation}
M_{DM}\left(R\right)=\rho_{0}\pi R_{0}^{3}\left\{ \log\left(\frac{R^{2}+R_{0}^{2}}{R_0^{2}}\right)+2\log\left(\frac{R+R_{0}}{R_{0}}\right)-2\arctan\left(\frac{R}{R_{0}}\right)\right\}. \label{eq:12}\end{equation}

For the baryons in the bulge, we use an exponential profile \cite{Gnedin:2004cx}
of the form \begin{equation}
\rho_{b}\left(R\right)=\frac{M_{bulge}}{8\pi R_{b}^{3}}e^{-R/R_{b}},\label{eq:13}\end{equation}
where $M_{bulge}=10^{10}$ M$_{\odot}$ \cite{Mo:2010} is the mass
of the bulge. This gives the baryonic mass distribution in the galactic
bulge \begin{equation}
M_{b}\left(R\right)=M_{bulge}\left\{ 1-e^{-R/R_{b}}\left(1+\frac{R}{R_{b}}+\frac{R^{2}}{R_{b}^{2}}\right)\right\} \label{eq:14}\end{equation}

We assume a Maxwell-Boltzmann velocity distribution for the dark matter particles
of the galactic halo, with a velocity dispersion $u\left(R\right)$
and a cutoff at the galactic escape velocity $v_{esc}\left(R\right)$:
\begin{equation}
f\left(R,\vec{v}_{h}\right)={1\over C\left(R\right)}e^{-v_{h}^{2}/u^{2}(R)}\label{eq:15}\end{equation}
 where $\vec{v}_{h}$ is the velocity of the dark matter particles
in the frame of the halo and
\newcommand{\erf}{\mathrm{erf}} $C(R)=\pi u^{2}\left(\sqrt{\pi}u\, \erf(v_{esc}/u)-2v_{esc}e^{-v_{esc}^{2}/u^{2}}\right)$
is a normalization constant such that $\int_{0}^{v_{esc}(R)}f\left(R,\vec{v}_{h}\right)d\vec{v}_{h}=1$.

The radial dependence of the velocity dispersion is obtained via the
virial theorem: \begin{equation}
u\left(R\right)=\sqrt{\frac{GM_{tot}\left(R\right)}{R}}\label{eq:16}\end{equation}
 where $M_{tot}=M_{DM}+M_{b}$, while $v_{esc}=\sqrt{2}u$.

Using the velocity distribution \eqref{eq:15}, going to center-of-mass
and relative velocities $\vec{v}_{CM}$ and $\vec{v}$ and performing
the integrals over $\vec{v}_{CM}$, we obtain for the mean pair-production
cross section times relative velocity: \begin{equation}
\left\langle \sigma_{ee}v\right\rangle =\frac{1}{u^{2}}\frac{\sqrt{2\pi}u\, \erf\left(\sqrt{2}v_{esc}/u\right)-4v_{esc}e^{-2v_{esc}^{2}/u^{2}}}{\left(\sqrt{\pi}u\, \erf\left(v_{esc}/u\right)-2v_{esc}e^{-v_{esc}^{2}/u^{2}}\right)^{2}}\int_{0}^{2v_{esc}}\sigma_{ee}\left(v\right)v^{3}e^{-v^{2}/2u^{2}}dv,\label{eq:17}\end{equation}
 which is also a function of $R$ through $u$ and $v_{esc}$. Putting
\eqref{eq:9}, \eqref{eq:11}, \eqref{eq:12}, \eqref{eq:14}, \eqref{eq:16}
and \eqref{eq:17} together allows us to compute the pair production
rate in the galactic bulge defined in \eqref{eq:10} as a function
of $\rho_{0}$ and $M$.
\begin{figure}[h]
    \begin{center}
        \includegraphics[scale=1.0]{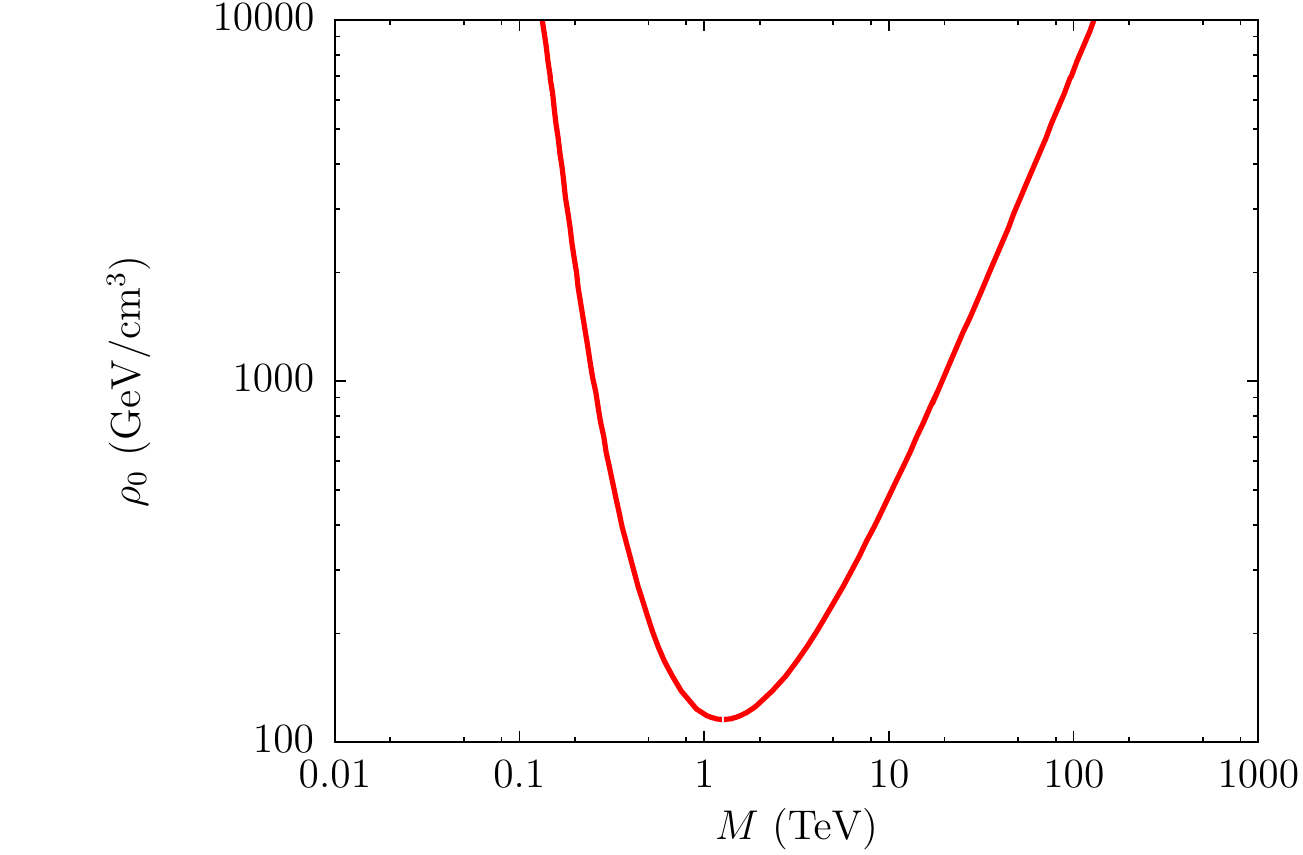}
        \caption{Values of the central dark matter density $\rho_{0}$ (GeV/cm$^{3}$)
and of the OHe mass $M$ (TeV) reproducing the excess of $e^{+} e^{-}$
pairs production in the galactic bulge. Below the red curve, the predicted rate is too low.}
\label{Flo:rho0_MOHe}
\end{center}
 %   \end{center}
\end{figure}
\section{Results}

The rate of excessive $e^{+} e^{-}$ pairs to be generated in the galactic bulge
was estimated in \cite{Finkbeiner:2007kk} to be ${\left. dN/dt\right|}_{obs}=3\times10^{42}$ s$^{-1}$
We computed ${\left. dN/dt\right|}_{ee}$ for a large range of
central dark-matter densities, going from $0.3$ GeV/cm$^{3}$ to
a ultimate upper limit of $10^{4}$ GeV/cm$^{3}$ \cite{Hernandez:2010sb}.
For each value of $\rho_{0}$, we searched for the mass $M$ of OHe
that reproduces the observed rate. The results are shown in Figure
\ref{Flo:rho0_MOHe}.

The observed rate can be reproduced from a value of $\rho_{0}\simeq 115$
GeV/cm$^{3}$, corresponding to an OHe mass of $M\simeq 1.25$ TeV.
As $\rho_{0}$ gets larger, two values of $M$ are possible, the lower
one going from $1.25$ TeV to $130$ GeV and the upper one going from
$1.25$ to $130$ TeV as $\rho_{0}$ goes from $115$ to $10^{4}$
GeV/cm$^{3}$.
\section{Conclusion}
The existence of heavy stable particles is one of the most popular solutions for the
dark matter problem.
Usually they are considered to be electrically neutral. But  dark matter can
potentially be made of
stable heavy charged particles bound in neutral atom-like states by Coulomb attraction.
An analysis of the cosmological data and of the atomic composition of the Universe
forces the particle to have charge $-2$.
$O^{--}$ is then trapped by primordial helium
in neutral O-helium states and this avoids the problem of overproduction of anomalous
isotopes, which are severely constrained by observations. Here we have shown that the
cosmological model of O-helium dark matter
can explain the puzzle of positron line emission from the center of our Galaxy.

The proposed explanation is based on the assumption that OHe dominates the dark-matter sector. Its collisions can lead to E0 de-excitations of the 2s states excited by the collisions. The estimated luminosity in the electron-positron annihilation line strongly depends not only on the mass of $O^{--}$, but also on the density profile and velocity distribution of dark matter in the galactic bulge. Note that the density profile we considered is used only to obtain a reasonable estimate for the uncertainties on the density in the bulge. It indeed underestimates the mass of the galaxy, but it shows that the uncertainties on the astrophysical parameters are large enough to reproduce the observed excess for a rather wide range of masses of $O^{--}$. For a fixed density profile and a fixed velocity distribution, only two values of the $O^{--}$ mass lead to the necessary rate of positron production. The lower value of this mass, which doesn't exceed $1.25$ TeV, is within the reach of experimental searches for multi charged stable heavy particles at the LHC.

\section*{Acknowledgments}
We express our gratitude to A.S. Romaniouk for discussions.

\end{document}